\documentclass[final,3p,times]{elsarticle}

\usepackage{amssymb,amsmath,amsfonts,mathrsfs,bm,gensymb}
\usepackage{graphicx,float,subfig}
\usepackage{lineno}

\usepackage{geometry}
\usepackage{hyperref}

\hypersetup{
    colorlinks=true,
    linkcolor=black,
    citecolor=black,
    filecolor=black,
    urlcolor=black,
}

\makeatletter
\def\ps@pprintTitle{%
   \let\@oddhead\@empty
   \let\@evenhead\@empty
   \let\@oddfoot\@empty
   \let\@evenfoot\@oddfoot
}
\makeatother

\biboptions{sort&compress}

\begin{document}

\begin{frontmatter}

\title{Compliant morphing structures from twisted bulk metallic glass ribbons}

\author[add1,add3]{P. Celli\fnref{eqc}\corref{corr1}}
\ead{paolo.celli@stonybrook.edu}
\author[add1]{A. Lamaro\fnref{eqc}}
\author[add1]{C. McMahan\fnref{eqc}}
\author[add2]{P. Bordeenithikasem}
\author[add2]{D. C. Hofmann}
\author[add1]{C. Daraio}
\fntext[eqc]{Equal contribution}
\cortext[corr1]{Corresponding author}

\address[add1]{Division of Engineering and Applied Science, California Institute of Technology, Pasadena, CA 91125, USA}
\address[add2]{Jet Propulsion Laboratory/California Institute of Technology, Pasadena, CA 91109, USA}
\address[add3]{Department of Civil Engineering, Stony Brook University, Stony Brook, NY 11794, USA}

\begin{abstract}
In this work, we investigate the use of pre-twisted metallic ribbons as building blocks for shape-changing structures. We manufacture these elements by twisting initially flat ribbons about their (lengthwise) centroidal axis into a helicoidal geometry, then thermoforming them to make this configuration a stress-free reference state. The helicoidal shape allows the ribbons to have preferred bending directions that vary throughout their length. These bending directions serve as compliant joints and enable several deployed and stowed configurations that are unachievable without pre-twist, provided that compaction does not induce material failure. We fabricate these ribbons using a bulk metallic glass (BMG), for its exceptional elasticity and thermoforming attributes. Combining numerical simulations, an analytical model based on {a geometrically nonlinear plate} theory and torsional experiments, we analyze the finite-twisting mechanics of various ribbon geometries. We find that, in ribbons with undulated edges, the twisting deformations can be better localized onto desired regions prior to thermoforming. Finally, we join multiple ribbons to create deployable systems { with complex morphing attributes enabled by the intrinsic chirality of our twisted structural elements}. Our work proposes a framework for creating fully metallic, yet compliant structures that may find application as elements for space structures and compliant robots.

\vspace{15px}
\noindent{\textbf{This article may be downloaded for personal use only. Any other use requires prior permission of the authors and Elsevier Publishing. This article appeared in}: \emph{Journal of the Mechanics and Physics of Solids} 145, 104129 (2020) \textbf{and may be found at}: \url{https://doi.org/10.1016/j.jmps.2020.104129}}
\vspace{10px}
\end{abstract}


\end{frontmatter}


\section{Introduction}
\label{s:intro}

Shape-changing structures are mechanical systems designed to undergo predictable changes of shape when subjected to external or internal stimuli. Typically, such structures are made of separate elements that can move relative to each other and are connected via kinematic joints, and the acts of deployment and retraction do not require the various elements to be dismounted~\cite{You1997, Pellegrino2001, You2012, Fenci2017}. They find use as everyday objects (e.g., foldable chairs and expandable toys), architectural elements (e.g., retractable roofs and pop-up domes), space structures (e.g., deployable booms, solar sails and starshades) and medical devices (e.g., stents and capsules for drug delivery). In space systems, deployable structures are necessary to satisfy increasingly stringent packaging ratios and weight requirements imposed by cubesats. One way to reduce weight and complexity in deployable systems is to replace multiple jointed parts with continua featuring compliant hinges. Here, we call these systems ``compliant morphing structures''. Examples of compliant structure classes are: origami, which feature axially-rigid but potentially-flexible panels connected by foldable creases~\cite{Guest1992, Schenk2013, Filipov2015}; kirigami, where creases are combined with cuts to expand the range of achievable morphed shapes~\cite{Sussman2015, Neville2016, Rafsanjani2017}; compliant mechanism-like structures, where bulky components are connected via thin flexures~\cite{Howell2001, Ion2016, Konakovic2016, Overvelde2017, Shang2018, Konakovic2018, Celli2018, Kamrava2019}; and creaseless foldable shell structures such as tape springs and slotted cylinders~\cite{Schleicher2015, Seffen1999, Walker2019, Sakovsky2019}. Through careful design, some of these compliant systems achieve extreme changes of shape that are typically unattainable with other strategies. Examples are systems that transform from flat configurations into 3D shapes~\cite{Dudte2016, Callens2017, Konakovic2018, Celli2018, Boley2019, Guseinov2020} and compact objects that deploy into large surfaces~\cite{Arya2017}.

In compliant structures, high stresses are typically concentrated at the creases/flexures. This makes it challenging to design low-part-count systems that possess complex and reversible morphing attributes and are simultaneously made of materials that provide the load-bearing capacity or durability required by certain structural applications. An attempt in this direction is the relization of additively manufactured, metallic origami~\cite{Harris2019}. Another example is represented by bulk metallic glass kirigami sheets~\cite{Chen2019}. Other researchers have attempted to use composites to create morphing systems~\cite{Neville2016, Deleo2020}. These structures typically require the union of multiple elements via pin-joints to achieve complex morphing scenarios~\cite{McHale2019} due to limitations in fabrication processes~\cite{Knott2019}. Others have considered origami systems with more complex compliant hinge geometries to reduce stresses~\cite{Delimont2015, Nelson2019}. 

To create compliant morphing structures made of materials relevant to structural engineering, systems that feature extremely compliant, yet robust and manufacturable hinges are needed. In recent years, ribbons (slender structural elements where length $\gg$ width $\gg$ thickness) have emerged as building blocks for morphing structures, as they can be bent and buckled~\cite{Morigaki2016, Gillman2018, Walker2019}, twisted~\cite{Mockensturm2000, Zhao2006, Cranford2011, Korte2011, Kit2012, Chopin2013, Armon2014, Dias2015, Boddapati2020} and sheared~\cite{Yu2019}. Their dimensions can be tailored to avoid the onset of plasticity when deformed. For example, sheets with ribbon-like features made of various materials (including metals) can be transformed into 3D objects via compressive buckling when triggered by the release of a pre-stretched substrate~\cite{Zhang2015, Liu2016, Shi2017, Fu2018, Guo2018}. The main issue with this approach in a structural setting is its limited scalability and its reliance on a substrate for deployment. The structural capacity of ribbon-based compliant systems can be improved when ribbons are joined and used as building blocks for free-standing structures, but few efforts have been made in this direction~\cite{Lipton2018, Zhang2019, Aza2019, Nguyen2019, McHale2019}. One constraint is the fact that a ribbon can only be significantly compacted by bending it about the axis aligned with the ``width'' direction, thus limiting the stowing configurations of ribbon-based structures.

\begin{figure}[!htb]
\centering
\includegraphics[scale=1.1]{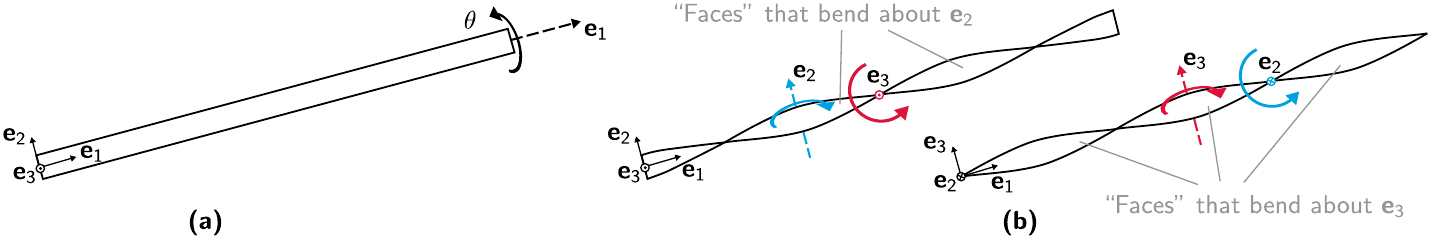}
\caption{Twisting ribbons to create structural elements with an expanded set of bending axes. (a) Original ribbon configuration, {where $\{\textbf{e}_i\}$ are the orthonormal basis vectors of a coordinate system aligned with the centerline of the ribbon.} (b) Ribbon configuration after a $\theta=3\pi$-degree twist about $\textbf{e}_1$, viewed from two different directions. In a twisted configuration, we call ``faces'' those regions that can bend about $\textbf{e}_2$ or $\textbf{e}_3$.}
\label{f:idea}
\end{figure}

In this work, we propose the combination of a design framework and a material choice to create ribbon-based compliant morphing structures. Our fundamental structural element is a pre-twisted bulk metallic glass (BMG) ribbon. Applying finite twists to the ribbon sketched in Fig.~\ref{f:idea}(a) about its longitudinal axis produces the beam-like structural element in Fig.~\ref{f:idea}(b). If we construct a fixed coordinate system with orthonormal basis vectors $\{\textbf{e}_i\}$ and align the twist axis of the ribbon with $\textbf{e}_1$, some ribbon cross-sections have preferred bending directions about $\textbf{e}_2$ and others about $\textbf{e}_3$. We will refer to these regions as ``faces'' throughout this article.
This expanded set of bending axes and the inherent chirality imparted via twisting allows for extreme compaction of the ribbon. 
BMGs have attractive properties for compliant structures~\cite{Homer2014, Chen2019} due to a broad elastic range, up to 2\% strain~\cite{Telford2004, Ashby2006, Kruzic2016}. Additionally, BMGs can be made into complex, stress-free geometries via thermoforming, where the alloy is heated above its glass transition temperature, reshaped, and quenched to avoid crystallization~\cite{Schroers2010}. In this work, we choose $\mathrm{Zr_{65}Cu_{17.5}Ni_{10}Al_{7.5}}$ BMG~\cite{Inoue1993} since it is widely studied in the literature and is commercially available in melt-spun ribbon form~\cite{Tent2018}. First, we provide a complete mechanistic analysis of the twisting process, and propose a ribbon configuration with undulated edges that allows us to localize the majority of the twist onto desired regions. The influence of various design parameters is analyzed via finite-element (FE) simulations and through an analytical model based on { a geometrically nonlinear plate} theory. We compare these results with torsion experiments on BMG ribbons. Once we have a complete mechanistic description of twisting, we thermoform ribbons into twisted shapes, and assemble them into structural prototypes of deployable mechanical systems, such as collapsible rings, spheres {and auxetic structures}. With this preliminary work we (i) shed light on the twisting mechanics of non-straight ribbons, (ii) illustrate the potential of twisted ribbons as structural elements for deployable structures, (iii) demonstrate that the combination of advanced materials such as BMGs and carefully-designed architectures can be leveraged to design compliant shape-morphing systems made of metals. Owing to the richness of achievable deformations we envision that, upon proper scaling, these structural systems could find application as components of deployable space structures (e.g., booms or rings for mesh-antennae) or as components for compliant, yet fully metallic robots.

In Section~\ref{s:design}, we illustrate the fundamental design parameters of our undulated ribbons and we provide an experimental characterization of the mechanical properties of BMG. In Section~\ref{s:twist}, we provide background information and results on the twisting mechanics of ribbons. We use FE simulations to understand the influence of the various design parameters and compare it to twisting experiments we conducted on BMG ribbons. We then adapt an analytical model introduced by Mockensturm to our case of undulated ribbons, and use it to analyze the influence of the design parameters on their elastic response. In Section~\ref{s:struct}, we describe our thermoforming setup and show that it enables the fabrication of pre-twisted ribbons that are subsequently spot-welded into structures that display extreme morphing capacity. Conclusions and future outlook are reported in Section~\ref{s:concl}.

\section{Ribbon design and material characterization}
\label{s:design}

\begin{figure}[!htb]
\centering
\includegraphics[scale=1.1]{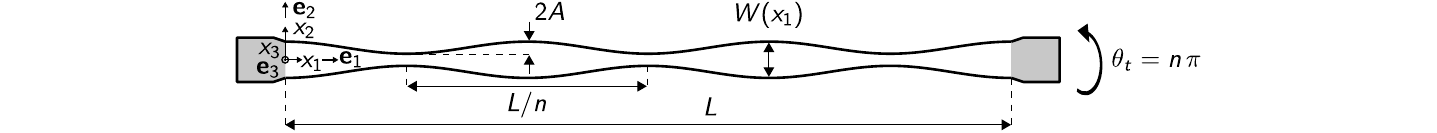}
\caption{Undulated ribbon characterized by thin necks and wide faces, with all the relevant geometrical parameters. The gray extremities/tabs of the ribbon are not part of the model, but facilitate clamping of the fabricated specimens. $\theta_t$ is the target twisting angle to align all necks with the $\textbf{e}_1$--$\textbf{e}_3$ plane and all faces with the $\textbf{e}_1$--$\textbf{e}_2$ plane.}
\label{f:geom}
\end{figure}

One of the ribbon geometries used in our work is shown in Fig.~\ref{f:geom}. As in Fig.~\ref{f:idea}, we define an orthonormal vector basis $\{\textbf{e}_i\}$, with material coordinates $x_i$.
The ribbon has thickness $H$ and its length is $L=180\,\mathrm{mm}$, unless otherwise specified (excluding the shaded gray tabs used for clamping purposes). The long edges of the ribbon have a sinusoidal profile with amplitude $A$ and wavelength $L/n$, where $n$ is the number of necks. The width of the ribbon follows the function $W(x_1)=w+2A[\cos(2\pi n x_1/L)-1]$, where $w=9\,\mathrm{mm}$ unless otherwise specified. In order to achieve a twisted state where all necks represent faces that can bend about the $\textbf{e}_3$ axis, an undulated ribbon needs to be twisted through a target angle $\theta_t=n\,\pi$.

\begin{figure}[!htb]
\centering
\includegraphics[scale=1.1]{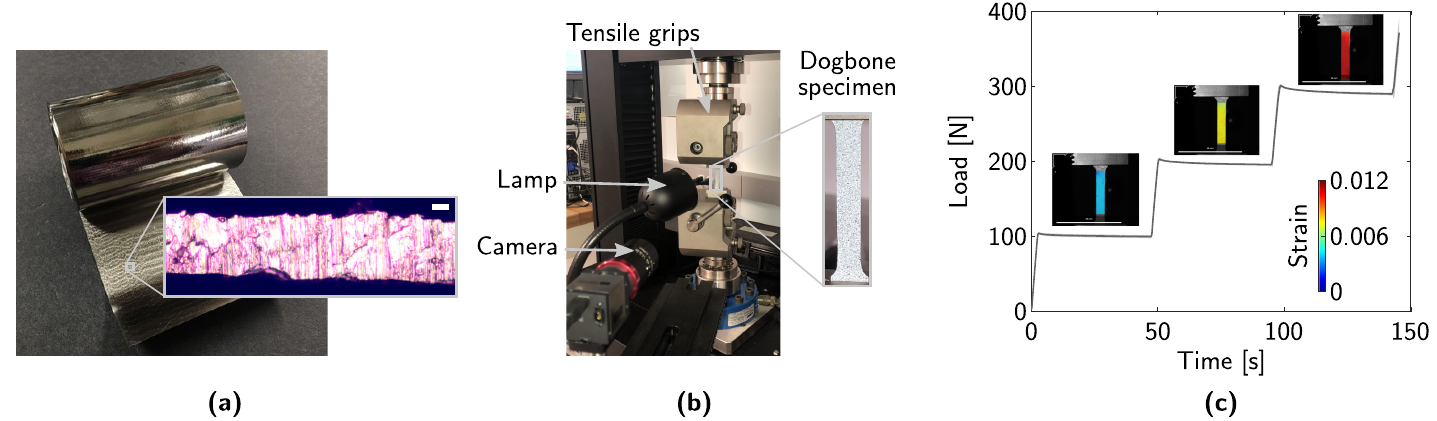}
\caption{Material characterization. (a) BMG roll ($\mathrm{Zr_{65}Cu_{17.5}Ni_{10}Al_{7.5}}$). The micrograph shows the irregular cross section of the roll (Scale bar: $10\,\mathrm{\mu m}$). (b) Tensile test setup to characterize the BMG sheets. (c) Load-time curve indicating our testing procedure; the specimen is pulled and the force is held constant at various force values to record images for the DIC procedure. The insets show the DIC-computed axial strain field.}
\label{f:mater}
\end{figure}

All ribbons in our work are manufactured from a melt-spun roll of the $\mathrm{Zr_{65}Cu_{17.5}Ni_{10}Al_{7.5}}$ alloy. The roll and a micrographic image showing the melt-spinning-induced irregularities of the cross-section are shown in Fig.~\ref{f:mater}(a).
For modeling purposes, we cut several BMG pieces from the same roll and measure their thickness using a microscope, finding an average thickness of $54\,\mathrm{\mu m}$. All specimens used in this work are obtained by creating drawings in MATLAB, cutting PETG masks with a Silhouette Cameo cutter, using these masks to mark the edges of the ribbons on the BMG roll, and manually cutting the roll. Since the mechanical performance of our ribbons is affected by the cross-sectional imperfections visible in Fig.~\ref{f:mater}(a), we measure the mechanical properties of several dogbone-shaped specimens. We do so with the universal testing machine shown in Fig.~\ref{f:mater}(b) (ADMET eXpert 8612 Table-Top Axial Torsion Test System, with a $25\,\mathrm{kN}$ axial load cell), equipped with grippers for tension tests. Our setup also features a high-definition camera (Edmund Optics EO-5023M) to record photographs that are analyzed via 2D digital image correlation (DIC). First, we perform a tensile test to understand the behavior of the material and to identify the limits of the linear elastic regime ({see ~\ref{a:mat}). We realize that the material behaves linear-elastically up until the breaking point, which occurs at a breaking strain $\varepsilon_b\approx1.7\%$ and at a stress $\sigma_b\approx1.2\,\mathrm{GPa}$}. In light of this, we test three specimens of equal dimensions following the load path illustrated in Fig.~\ref{f:mater}(c). A specimen is pulled up to loads of 100, 200 and 300 N. At those values, the force is kept constant for 45 seconds to allow us to record a picture of the specimen. We compare these images to the undeformed configuration using the DIC software nCorr~\cite{Blaber2015} to extract the strains in the plane of the specimen. As a result, we can measure Young's modulus $E$ and Poisson's ratio $\nu$ for the material by averaging these quantities across specimens. We obtain the following values: $E=78\,\mathrm{GPa}$ and $\nu=0.355$.

\section{Twisting mechanics}
\label{s:twist}

This Section is dedicated to the analysis of the twisting mechanics of undulated ribbons, with the goal of understanding what geometries yield BMG ribbons that can be twisted into desired shapes, where the bending axes of adjacent faces are perpendicular to each other as sketched in Fig.~\ref{f:idea}(b). 

For some boundary conditions, it is observed that twisting a ribbon leads to mechanical instabilities that result in the appearance of wrinkle-like patterns. This behavior was first observed by A.\ E.\ Green in 1936~\cite{Green1936, Green1937}, and has received renewed attention since the early 2000's. Mockensturm's work on the topic is the most general from a modeling standpoint, where a fully-nonlinear plate theory is used to model the twisting behavior and elastic instabilities of arbitrarily-wide ribbons subjected to large twists~\cite{Mockensturm1998, Mockensturm1999, Mockensturm2000}. A very comprehensive article on this phenomenon by Chopin and Kudrolli~\cite{Chopin2013} used experiments and scaling arguments to map various buckling modes of twisted, pre-stretched ribbons clamped at their edges. Their conclusions are that there is a critical pre-stretch at which there is a transition between lateral and longitudinal buckling modes, and that the ribbon geometry strongly influences the critical twists and achievable post-buckled shapes. 

In our work, we are interested in avoiding these instabilities. In fact, for our ribbons to morph into structural elements with multiple preferred bending axes, we need {to avoid any localization of curvature} that would compromise their deployability and their post-twisting response. However, special considerations need to be made since our ribbons feature non-straight edges---a scenario that is seldom considered in the existing literature~\cite{Dias2015}. Therefore, we use numerical simulations to predict the principal strains and the deformed shapes achievable by twisting undulated ribbons. These simulations are validated via torsional experiments on BMG specimens. To gain a better understanding of the mechanics involved and of the influence of the design parameters, we adapt the model developed by Mockensturm~\cite{Mockensturm2000} to the case of ribbons with non-constant cross-sections. 

\subsection{Numerical modeling}
\label{s:num}

Our numerical, finite element (FE) simulations are conducted using the commercial software Abaqus. The ribbon configurations we consider have the dimensions reported in Section~\ref{s:design}, and varying numbers of necks $n$ and undulation amplitudes $A$. We consider four-node reduced-integration shell elements (of the S4R type) with 7 through-the-thickness integration points. These elements are suitable for geometrically-nonlinear analyses. The material response is considered to be linear over finite strains, an assumption that is acceptable for a material like BMG (see \ref{a:mat}). { Each ribbon is clamped at the bottom edge, while all nodes of the top edge are fixed to a fictitious reference point where we apply the load}. The solution is carried out in two separate steps. First, we use an implicit/static analysis to model the axial pre-stretching step necessary to avoid longitudinal instabilities. This is enforced by applying an initial displacement of $0.1\,\mathrm{mm}$ between clamped boundaries. Then, we use an explicit/dynamic analysis to model the twisting process. This is done in order to speed up the computation time with respect to the standard implicit solver. We use a mass-scaling approach to accelerate computations, where the density of the material is artificially scaled to increase the stable time increment. To ensure that the model reflects the quasi-static nature of the twisting process, we monitor the total kinetic energy of our system and make sure it remains below 5$\%$ of the total energy in each simulation. 

\begin{figure}[!htb]
\centering
\includegraphics[scale=1.1]{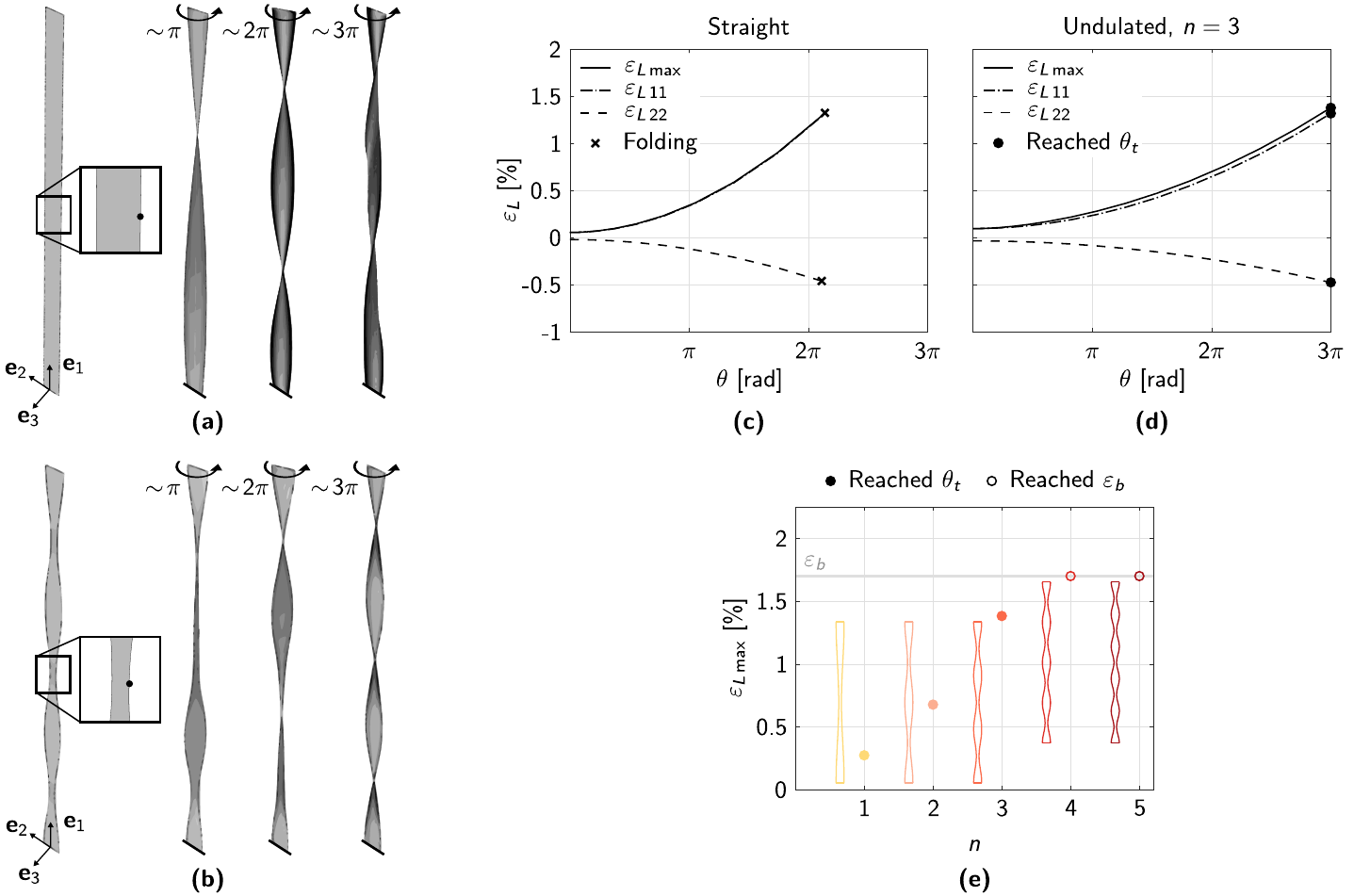}
\caption{Numerical (FE) results on twisting. (a) Initial and deformed configurations for an initially-straight ribbon subjected to pre-stretch and torsion. Darker colors indicate regions of higher maximum principal strain, {and serve the sole purpose of qualitatively showing where the strains are largest.} (b) Same as (a), but for an undulated ribbon with $n=3$ and $A=w/6$. The circular markers in the undeformed configurations indicate the locations where strains are the largest. (c) Logarithmic strain versus twist angle for a straight ribbon. The crosses indicate when the ribbon reached an unwanted self-folded configuration. (d) Logarithmic strain versus twist angle for an undulated ribbon with $n=3$ and $A=w/6$. (e) Effects of the number of necks $n$ on the maximum strain, with $A=w/6$ fixed. Recall that the breaking strain for this material is $\varepsilon_b=1.7\%$. 
}
\label{f:numstr}
\end{figure}

First, we analyze the response of a straight ribbon; this is summarized in Fig.~\ref{f:numstr}(a,c).
In Fig.~\ref{f:numstr}(a), we illustrate the undeformed configuration. The ribbon is clamped at both bottom and top ends. The load is modeled as a displacement along $\textbf{e}_1$ followed by twist about $\textbf{e}_1$ applied to the top end of the ribbon. This figure also illustrates the ribbon for various levels of twisting. Darker colors indicate regions of higher maximum principal strain. One can qualitatively see that, as already known from the literature~\cite{Chopin2013}, larger strains concentrate at the edges of the ribbon. Thus, we extract quantitative information on the response at the critical point illustrated in the inset (located at $x_1=L/2$ and $x_2=-W(x_1)/2$ in the undeformed configuration). Considering the total twisting angle $\theta$ between the top and bottom edges as our variable, we monitor the evolution of the maximum principal ($\varepsilon_{L\,\mathrm{max}}$), axial ($\varepsilon_{L\,11}$) and lateral ($\varepsilon_{L\,22}$) strains and plot them in Fig.~\ref{f:numstr}(c). All strains are logarithmic. We can see that the maximum principal strain coincides with the axial strain, and that they are both nonzero at $\theta=0$ due to the pre-stretch. On the other hand, the lateral strain is compressive. This behavior is due to Poisson's effects that balance the twisting-induced tension, and it is known to lead to lateral buckling~\cite{Chopin2013}. This behavior is actually visible in our numerical results in the form of a self-folding that takes place after $\theta=2\pi$ (see also the deformed shape at $3\pi$ in Fig.~\ref{f:numstr}(a)). After self-folding occurs, the strains assume values that are strongly dependent on the assumed contact parameters, and are therefore deemed unrealistic.

In Fig.~\ref{f:numstr}(b,d), we report the response of an undulated ribbon with $n=3$ necks and undulation amplitude $A=w/6$. In this case, the maximum strains are achieved at the edge of a neck region. The strain plot illustrates that the maximum strains achieved are less than those in the straight ribbon and remain below the breaking strain of 1.7$\%$. Moreover, no self-folding is observed prior to the target angle $\theta_t=3\pi$ due to smaller lateral compressive strains. The final twisted configuration is illustrated in Fig.~\ref{f:numstr}(b) and, as expected, it features necks parallel to the $\textbf{e}_1$--$\textbf{e}_3$ plane and wide faces parallel to $\textbf{e}_1$--$\textbf{e}_2$. 

{Now that we have illustrated the benefits of the undulated edge geometry, we use our numerical model to analyze the effects of the number of necks $n$ on the twisting response. A more detailed parametric analysis is then carried out analytically in Section~\ref{s:analy}. In Fig.~\ref{f:numstr}(e), the markers indicate the maximum principal strain as a function of the number of necks. Each value is recorded at a target twist angle that is dependent on the number of necks. We also superimpose the color-coded silhouettes of the ribbons as a visual aid. Increasing the number of necks causes the maximum principal strain to increase. For the 4 and 5 neck cases, we reach the breaking strain before reaching the target angles of $4\pi$ and $5\pi$, respectively. This indicates that, for the ribbon dimensions we selected, more than 3 necks (wavelengths less than $6\,\mathrm{cm}$) are not admissible.}

\subsection{Experimental validation}
\label{s:exp}

\begin{figure}[!htb]
\centering
\includegraphics[scale=1.1]{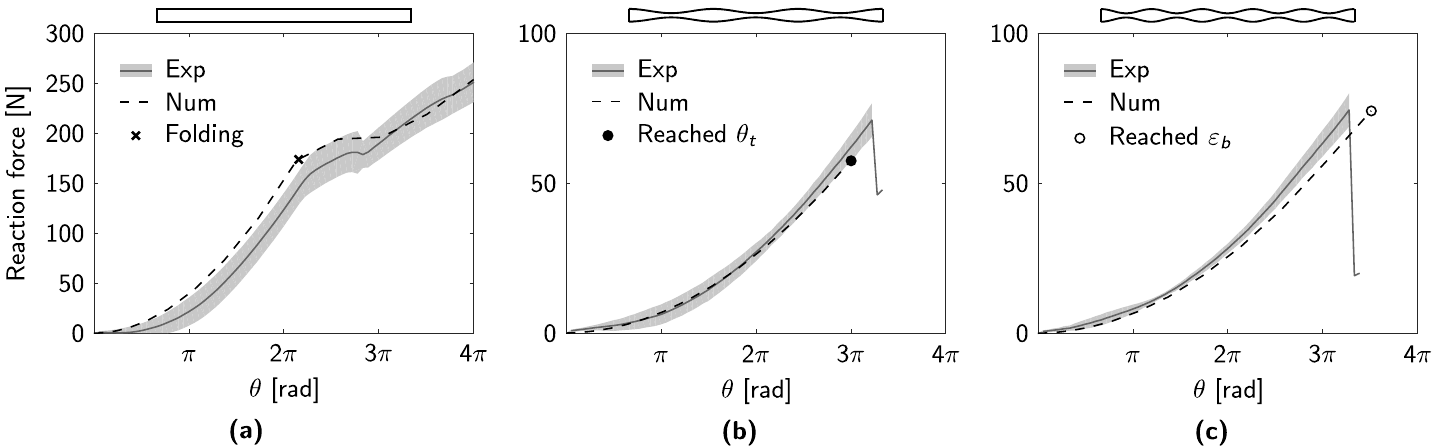}
\caption{Experimental validation of the numerical predictions on twisting. (a) Axial reaction force versus twist angle for a straight ribbon. The dark gray line is the mean and the shaded light gray area indicates the standard deviation of measurements performed on three specimens. The cross marker indicates the $\theta$ angle at which the simulation indicates self-folding. (b), (c) Same as (a), but for a 3-neck and 5-neck ribbon, respectively. Both cases feature $A=w/6$. A sharp drop in the experimental curve indicates failure.}
\label{f:exp}
\end{figure}

To experimentally validate our numerical predictions, we perform torsional tests on ribbons of various geometries. These tests are carried out using the same apparatus we used for the axial experiments described in Section~\ref{s:design}, using grippers designed for torsion. The comparison between experimental and numerical results for three ribbon geometries is shown in Fig.~\ref{f:exp}.
In all cases, we plot the axial reaction force developed during twisting, as a function of the twisting angle. For the straight ribbon configuration, shown in Fig.~\ref{f:exp}(a), we can see that the numerical response follows the experimental trend both during the monotonic force increase that is observed before self-folding and during the non-monotonic regime that occurs after the ribbon self-folds. The self-folding point achieved numerically is indicated by the cross marker. Despite the incidence of self-folding, the experiments illustrate that the ribbon does not fail in the 0-4$\pi$ twist range. When considering a ribbon with $n=3$ and $A=w/6$, as illustrated in Fig.~\ref{f:exp}(b), we can see that numerics and experiments agree well. Moreover, as predicted in Section~\ref{s:num}, the experiments confirm that this ribbon does not fail at the target angle of 3$\pi$. Finally, in Fig.~\ref{f:exp}(c), we confirm the numerical prediction that a ribbon with $n=5$ and $A=w/6$ fails (reaching the breaking strain) long before reaching the target angle 5$\pi$. 

These experiments serve as a partial validation of our numerical model. They provide insight into the axial response of the ribbons, but offer no information on the lateral stresses that arise during twisting. Since twisting induces out of plane deformations, we cannot reliably use 2D DIC; moreover, 3D DIC would only be useful for small twists. For these reasons, we develop an analytical model to verify the numerical prediction that ribbons with undulated edge geometries experience considerably smaller compressive stresses in the lateral direction, thus delaying the onset of buckling due to twist.

\subsection{Analytical modeling}
\label{s:analy}
The model we employ to analyze the state of strain in the ribbons prior to thermoforming is heavily based on a fully nonlinear, {geometrically exact description of rectangular ribbons} that was developed by Mockensturm~\cite{Mockensturm2000}. We re-derive this model while making minor modifications in order to extend it to ribbons with edges that are symmetric, but not straight. We provide an overview of the model and its assumptions, compare the analytical and computational predictions, {and perform a more extensive parametric analysis on undulated ribbons, while highlighting some limitations of the approach.}

\subsubsection{Reference and  deformed configurations}
\label{s:conf}
A ribbon's material particle positions in an untwisted reference configuration are $\mathbf{X}=x_\alpha \mathbf{e}_\alpha+x_3 \mathbf{e}_3$. {Throughout this text, the indices $\alpha$ and $\beta$ pertain to the mid-surface of the ribbon, the index `3' corresponds to the direction normal to the surface, and we use the Einstein summation convention for repeated indices.} The coordinates $x_i$ are convected and material, and $0<x_1<L,\ |x_2|<W(x_1)/2,\ |x_3|<H/2$ for a ribbon of uniform length $L$ and thickness $H$, and varying width $W(x_1)$. Our fixed coordinate frame $\{\mathbf{e}_i\}$ is orthonormal. The mapping $\mathbf{\boldsymbol{\chi}^*}(x_i)=\boldsymbol{\chi}(x_\alpha)+x_3\mathbf{\hat{v}}^3$ describes these particles in the twisted configuration, where $\boldsymbol{\chi}(x_\alpha)$ is the deformation mapping of the midplane of the ribbon, and $\mathbf{\hat{v}}^3$ is the outward unit normal to the surface $\mathcal{S}$ defined by $\mathbf{\boldsymbol{\chi}}(x_\alpha)$. It is assumed that unit normals to the surface $x_\alpha \mathbf{e}_\alpha$ are mapped to unit normals of $\mathcal{S}$, with vanishing transverse strains. This is an assumption that is valid within the thin { plate} approximation framework {and is called the Love-Kirchhoff hypothesis.}

After the ribbon is subjected to a pre-stretch and torsion, its mid-surface assumes a helicoidal geometry, where the expression for $\boldsymbol{\chi}(x_\alpha)$ is shown in Eq.~\ref{defsurf} below. For a cross-section located at $x_1$, $f(x_1,x_2)$ is the mapping of the particles in the lateral direction $x_2$, $\lambda_1(x_1)$ is the local axial pre-stretch, and $\theta(x_1)$ is the local twist angle relative to the supported edge at $x_1=0$. {Throughout this text, when the material coordinate is not specified, we are evaluating the twist angle function at the end of the ribbon. Namely, $\theta\equiv\theta(L)$.}
\begin{equation}\label{defsurf}
\mathbf{\boldsymbol{\chi}}(x_\alpha) = \lambda_1(x_1) x_1 \mathbf{e}_1+f(x_1,x_2)\cos(\theta(x_1))\mathbf{e}_2+f(x_1,x_2)\sin(\theta(x_1))\mathbf{e}_3.
\end{equation}
The expression above is slightly different from what is used by Mockensturm, as it accounts for variations in $\lambda_1,\ f$, and $\theta'$ as a function of axial position $x_1$. These functions are calculated in the following subsection.

\subsubsection{Pre-stretch and twist as a function of axial position}
\label{s:pretw}

Static equilibrium implies that the total axial force
$F$ acting on each cross-section is independent of position $x_1$. Therefore, the local linear axial strain $du_1/dx_1$ due to pre-stretch can be written as:
\begin{equation}
    \frac{du_1 (x_1)}{dx_1}=\frac{F}{EHW(x_1)}.
\end{equation}
The total stretch $\lambda_{tot}$ of the ribbon is obtained by integration:
\begin{equation}
\lambda_{tot}=\frac{1}{L}\Bigg(L+\int_0^L \frac{du_1}{dx_1}dx_1\Bigg)=1+\frac{F}{EHL}\int^L_0\frac{1}{W(x_1)}dx_1.
\end{equation}
This total pre-stretch is prescribed in our experiments, and therefore we know the total force $F$ that is being exerted at all ribbon cross-sections:
\begin{equation}
F=\frac{EHL(\lambda_{tot}-1)}{\int_{0}^{L}\frac{1}{W(x_1)}dx_1}.
\end{equation}
Knowing $F$, the local axial pre-stretch $\lambda_1(x_1)$ of an infinitesimally long cross-section at $x_1$ is obtained from the previously calculated quantities:
\begin{equation}\label{prestr}
\lambda_1(x_1)=1 + \frac{du_1 (x_1)}{dx_1}=1+\frac{L(\lambda_{tot}-1)}{W(x_1)\int_{0}^{L}\frac{1}{W(x_1)}dx_1}.
\end{equation}

Recalling the assumption that unit normals to the reference surface are mapped to unit normals of the deformed surface and that transverse strains vanish, we calculate the twist of the ribbon as a function of $x_1$. We note that the shear modulus $G$ is a constant and that the torque $T$ is the same at all ribbon cross-sections, so the twist rate of a ribbon is given by:
\begin{equation}
\theta'(x_1)=\frac{T}{GJ(x_1)}.
\end{equation}
The polar moment of inertia for a slender rectangular cross section ($W\gg H$) is $J=WH^3/3$. Since we prescribe the total twist of the ribbon's supported edges $\theta(L)$, we can calculate the ratio $T/G$ (for brevity, we omit the algebraic steps that are similar to the calculation of $F$ above) and thus know the twist rate as a function of axial position $x_1$:
\begin{equation} 
\theta'(x_1)=\frac{\theta(L)}{W(x_1)}\frac{1}{\int_{0}^{L}\frac{1}{W(x_1')}dx_1'}.
\end{equation}
We integrate this expression to find that the twist of the ribbon at $x_1$ relative to the fixed support at $x_1=0$ is:
\begin{equation}\label{twrate}
\theta(x_1)=\theta(L)\ \frac{\int_{0}^{x_1}\frac{1}{W(x_1')}dx_1'}{\int_{0}^{L}\frac{1}{W(x_1')}dx_1'}.
\end{equation}

\subsubsection{Kinematic measures}
\label{s:kin}

Now that expressions for $\lambda_1(x_1)$ and $\theta(x_1)$ are given in Eqs.~\ref{prestr} and \ref{twrate}, we must calculate $f(x_1,x_2)$ to complete our description of the deformed surface $\boldsymbol{\chi}$, represented by Eq.~\ref{defsurf}. Here, we follow Mockensturm's calculations closely but, unlike in his work, we account for the non-constant nature of $\lambda_1$ and $\theta'$. 

Each term that appears in the final elastic equilibrium equation is a function of the covariant and/or contravariant basis vectors of the ribbon's deformed configuration. The covariant basis on $\mathcal{S}$ is given by $\mathbf{v}_{\alpha}=\partial\mathbf{\boldsymbol{\chi}}/\partial x_\alpha$, and the reciprocal, contravariant basis $\mathbf{v}^{\alpha}$ to $\mathcal{S}$ is constructed such that $\mathbf{\hat{v}}_3\cdot\mathbf{v}_\alpha=0,\ \mathbf{\hat{v}}_3=\mathbf{\hat{v}}^3,\ \mathbf{v}^i \cdot \mathbf{v}_j=\delta^i_j.$ Here, $\delta^i_j$ is the Kronecker delta. {To simplify our calculations, we note that $\partial f(x_1,x_2)/\partial x_1$ is very small compared to $f(x_1,x_2)$ and $\partial f(x_1,x_2)/\partial x_2$ everywhere in the ribbon (this is supported by shear being negligible in our numerical simulations). We also can calculate that $x_1\lambda_1'(x_1)\ll\lambda_1(x_1)$ everywhere in the ribbon and that $x_1\theta''(x_1)\ll\theta'(x_1)$ in the regions surrounding the narrowest and widest cross-sections of the ribbon. In particular, $\lambda_1'(x_1)=\theta''(x_1)=\partial f(x_1,x_2)/\partial x_1=0$ where $W'(x_1)=0$ (the width extrema). Note that the numerical results tell us that stresses are global minima or maxima at these exact cross-sections. In the following, we keep our derivation general to a small region surrounding these width extrema, and we ignore these small terms in the expressions for the basis vectors $v_i$ and $v^i$.} Note that, given the above simplifications, we denote $f'(x_1,x_2)\equiv\partial f(x_1,x_2)/\partial x_2$ for the purpose of concise notation.

Our covariant basis vectors $\textbf{v}_i$ are: 
\begin{equation} 
\textbf{v}_1=\begin{Bmatrix}
\lambda_1(x_1)\\ -f(x_1,x_2)\theta'(x_1)\sin\Big(\theta(x_1)\Big)\\
f(x_1,x_2)\theta'(x_1)\cos\Big(\theta(x_1)\Big)
\end{Bmatrix},\ \ \ \textbf{v}_2=\begin{Bmatrix}
0\\ f'(x_1,x_2)\cos\Big(\theta(x_1)\Big)\\ f'(x_1,x_2)\sin\Big(\theta(x_1)\Big)\end{Bmatrix},
\end{equation}
\begin{equation*}
\textbf{v}_3=\frac{1}{\sqrt{\Big(\lambda_1(x_1)^2+\theta'(x_1)^2f(x_1,x_2)^2\Big)f'(x_1,x_2)}}\begin{Bmatrix}-\theta'(x_1)f(x_1,x_2)f'(x_1,x_2)\\
-\lambda_1(x_1)f'(x_1,x_2)\sin\Big(\theta(x_1)\Big)\\
\lambda_1(x_1)f'(x_1,x_2)\cos\Big(\theta(x_1)\Big)\end{Bmatrix}.
\end{equation*}
We then calculate the contravariant basis vectors:
\begin{equation}
\textbf{v}^1=\frac{1}{\lambda_1(x_1)^2+f(x_1,x_2)^2\theta'(x_1)^2}\begin{Bmatrix}\lambda_1(x_1)\\ -f(x_1,x_2)\theta'(x_1)\sin\Big(\theta(x_1)\Big)\\ f(x_2)\theta'(x_1)\cos\Big(\theta(x_1)\Big)\end{Bmatrix},\ \ \textbf{v}^2=\frac{1}{f'(x_1,x_2)}\begin{Bmatrix}0\\ \cos\Big(\theta(x_1)\Big)\\ \sin\Big(\theta(x_1)\Big)\end{Bmatrix},
\end{equation} 
\begin{equation*} 
\textbf{v}^3=\frac{1}{\sqrt{\Big(\lambda_1(x_1)^2+\theta'(x_1)^2f(x_1,x_2)^2\Big)f'(x_1,x_2)}}\begin{Bmatrix}-\theta'(x_1)f(x_1,x_2)f'(x_1,x_2)\\ -\lambda_1(x_1)f'(x_1,x_2)\sin\Big(\theta(x_1)\Big)\\
\lambda_1(x_1)f'(x_1,x_2)\cos\Big(\theta(x_1)\Big)\end{Bmatrix}.
\end{equation*}
The first and second fundamental forms $a_{\alpha\beta}$ and $b_{\alpha\beta}$, respectively, and the Christoffel symbols {of the second kind} $\Gamma_{ij}^k$ are used to provide local descriptions of $\mathcal{S}$, and are given by:
\begin{equation}\label{forms}
a_{\alpha\beta}=\mathbf{v}_\alpha\cdot\mathbf{v}_\beta,\ \ b_{\alpha\beta}=\Gamma^3_{\alpha\beta}=\mathbf{v}_{\alpha,\beta}\cdot\mathbf{\hat{v}}_{3},
\end{equation}
\begin{equation*}
\Gamma^\lambda_{\alpha\beta}=\mathbf{v}_{\alpha,\beta}\cdot \mathbf{v}^\lambda,\ \ \Gamma^\beta_{3\alpha}=-b_{\alpha}^{ \beta}=\mathbf{\hat{v}}_{3,\alpha}\cdot\mathbf{v}^{\beta},\ \ \Gamma^3_{3i}=\Gamma^i_{33}=0.
\end{equation*}
{In the definitions shown above (and throughout the remainder of the text), the underscore comma designates partial differentiation with respect to the corresponding coordinate component ($\mathbf{v}_{,i}\equiv \partial\mathbf{v}/\partial x_i$). We note that our tensors $b_{\alpha\beta}$ and $b_\alpha^\beta$ have identical matrix components. See \cite{Niordson1985} for a more thorough discussion of the relationship between the second fundamental form and Christoffel symbols of the second kind.} The matrix components of these forms are given below:
\begin{equation}\label{forms}
[a_{\alpha\beta}]=\begin{bmatrix}
\lambda_1(x_1)^2+f(x_1,x_2)^2\theta'(x_1)^2 & 0\\
0 & f'(x_1,x_2)^2
\end{bmatrix},
\end{equation}
\begin{equation*}
[b_{\alpha\beta}]=[\Gamma^3_{\alpha\beta}]=\frac{\lambda_1(x_1)\theta'(x_1)f'(x_1,x_2)}{\sqrt{\lambda_1(x_1)^2+\theta'(x_1)^2f(x_1,x_2)^2}}\begin{bmatrix} 0 &1\\1&0 \end{bmatrix},
\end{equation*}
\begin{equation*}
[\Gamma^1_{\alpha\beta}]=\frac{f(x_1,x_2)f'(x_1,x_2)\theta'(x_1)^2}{\lambda_1^2+f(x_1,x_2)^2\theta'(x_1)^2}\begin{bmatrix}
0&1\\1&0
\end{bmatrix},
\end{equation*}
\begin{equation*}
[\Gamma^2_{\alpha\beta}]=\frac{1}{f'(x_1,x_2)}\begin{bmatrix}
-f(x_1,x_2)\theta'(x_1)^2 &0\\0&f''(x_1,x_2)
\end{bmatrix},
\end{equation*}
\begin{equation*}
\Gamma^3_{3i}=\Gamma^i_{33}=0.
\end{equation*}
When the ribbons are mapped into the deformed configuration, the first fundamental form $a_{\alpha\beta}$ characterizes the in-plane stretches and {the second fundamental form $b_{\alpha\beta}$ describes the inner products between the partial derivatives of the covariant basis vectors and the unit normal, thus capturing out-of-plane bending. The mixed component form $b_{\alpha}^{\beta}$ defined in Eq.~\ref{forms} has the same matrix components as $b_{\alpha\beta}$ in our case and} captures the inner products between the partial derivatives of the local unit normal and the contravariant basis vectors, essentially describing the rotation of the unit normal as we move along the surface. The connection between these forms and the tensors for in-plane strain $\mathbf{C}$ and bending $\mathbf{\Lambda}$ commonly used in plate and shell mechanics is described more precisely below~\cite{Naghdi1982}:
\begin{equation}\label{straintensors}
\mathbf{C}=a_{\alpha\beta}(\mathbf{e}_\alpha\otimes\mathbf{e}_\beta),\ \ \mathbf{\Lambda}=-(b_{\alpha\beta}+b_{\beta\alpha})(\mathbf{e}_\alpha\otimes\mathbf{e}_\beta).
\end{equation}

\subsubsection{Material model and stress resultants}
\label{s:mat}

Mockensturm's usage of a Saint-Venant-Kirchhoff material model is also appropriate for our ribbons due to the large regime of elastic linearity displayed by the BMG. This model uses the following strain energy function:
\begin{equation} 
\varphi(\mathbf{C,\Lambda})=\frac{KH^2}{24}\bigg(\nu\frac{(\mathbf{\Lambda\cdot I})^2}{4}+(1-\nu)\frac{\mathbf{\Lambda\cdot\Lambda}}{4}\bigg)+\frac{K}{2}\bigg(\nu\frac{(\mathbf{C\cdot I}-3)^2}{4}+(1-\nu)\frac{(\mathbf{C-I})\cdot(\mathbf{C-I})}{4}\bigg)\ ,
\end{equation}
where $K=EH/(1-\nu^2)$, $E$ is Young's modulus, and $\nu$ is Poisson's ratio. The in-plane and bending stress resultants are:
\begin{equation} 
\mathbf{N}=2\frac{\partial\varphi(\mathbf{C,\Lambda})}{\partial \mathbf{C}},\ \ \ \ \ \ \ \mathbf{M}=2\frac{\partial\varphi(\mathbf{C,\Lambda})}{\partial \mathbf{\Lambda}}.
\end{equation}
In matrix form, these resultants are:
\begin{equation*}
\mathbf{N}=\frac{K}{2}\begin{bmatrix}
\lambda_1(x_1)^2-1+\theta'(x_1)^2f(x_1,x_2)^2+\nu\big(f'(x_1,x_2)^2-1\big)&0\\
0&\nu\big(\lambda_1(x_1)^2-1+\theta'(x_1)^2f(x_1,x_2)^2\big)+f'(x_1,x_2)^2-1
\end{bmatrix},
\end{equation*}
\begin{equation}\label{strres}
\mathbf{M}=-\frac{KH^2(1-\nu)\lambda_1(x_1)\theta'(x_1)f'(x_1,x_2)}{12\sqrt{\lambda_1(x_1)^2+\theta'(x_1)^2f(x_1,x_2)^2}}\begin{bmatrix}
0&1\\
1&0
\end{bmatrix}.
\end{equation}
We must now enforce our assumptions that transverse strains vanish and that unit normals to the reference configuration remain unit normals after deformation. The following constraint stress tensor is used to this purpose in this restricted kinematics plate theory and is calculated through the equilibrium equations in the next section:
\begin{equation} 
\mathbf{Q}=Q^\alpha(\mathbf{e}_\alpha\otimes\mathbf{e}_3+\mathbf{e}_3\otimes\mathbf{e}_\alpha)+Q^3\mathbf{e}_3\otimes\mathbf{e}_3.
\end{equation}

\subsubsection{Resolving the PDEs governing equilibrium onto the contravariant basis}
\label{s:pde}

The two equations describing equilibrium are derived fully in Mockensturm's doctoral dissertation \cite{Mockensturm1998} and are given below:
\begin{equation}\label{equil1}
[N^{\alpha\Gamma}\textbf{v}_\alpha+Q^\Gamma \mathbf{\hat{v}}_3+M^{\alpha\Gamma}\mathbf{\hat{v}}_{3,\alpha}]_{,\Gamma}=0,
\end{equation}
\begin{equation}\label{equil2}
[M^{\alpha\Gamma}\mathbf{v}_\alpha]_{,\Gamma}-[Q^\alpha\mathbf{v}_\alpha+Q^3\mathbf{\hat{v}}_3]=0.
\end{equation}
By resolving Eq.~\ref{equil2} onto $\mathbf{v}^\beta$ and $\mathbf{v}^3$ we obtain, respectively:
\begin{equation}
Q^\beta=M^{\beta\Gamma}_{,\Gamma}+M^{\alpha\Gamma}\Gamma^{\beta}_{\alpha\Gamma},
\end{equation}
\begin{equation*}
Q^3=M^{\alpha\Gamma}b_{\alpha\Gamma}.
\end{equation*}
By inserting $\mathbf{Q}$ into Eq.~\ref{equil1}, resolving the PDEs onto the contravariant basis $\textbf{v}^i$ and eliminating the zero-valued terms, we obtain the following statement of equilibrium:
\begin{multline}\label{equil3}
N^{11}_{,1}+N^{22}_{,2}+N^{11}\Gamma^2_{11}+N^{22}\Gamma^2_{22}+2(M^{12}_{,12}+M^{12}_{,1}\Gamma^1_{12}+M^{12}\Gamma^1_{12,1})+...\\-2(M^{12}_{,1}+M^{12}_{,2}+M^{12}\Gamma^1_{12})b^2_1-M^{12}(b^2_{1,1}+b^2_{1,2}+\Gamma^2_{11}b^2_1+\Gamma^2_{22}b^2_1)=0.
\end{multline}

The difference between our result at this point and what is shown in Mockensturm's work is the inclusion of terms where there are partial derivatives of the stress tensors in the $x_1$ direction due to the non-constant functions $\lambda_1(x_1)$ and $\theta'(x_1)$.

\subsubsection{Computing the lateral stretch using a perturbation method}
\label{s:pert}

{We proceed by stating that for this analytical model to be accurate, the amplitude of the edge undulations must be much smaller than the wavelength. This holds for most of our ribbons and we discuss the limitations of the model in Section~\ref{s:param}. In light of this consideration, we make a few simplifications driven by $W'(x_1)$ being small everywhere. We now focus our analysis at the width extrema (where $W'(x_1)=0$). At these specific cross-sections}, $N^{11}_{,1}=M^{12}_{,1}=M^{12}_{,12}=b^2_{1,1}=\Gamma^1_{12,1}=0$ and the equilibrium statement given by Eq.~\ref{equil3} becomes:
\begin{equation} 
N^{22}_{,2}+N^{11}\Gamma^2_{11}+N^{22}\Gamma^2_{22}-2(M^{12}_{,2}+M^{12}\Gamma^1_{12})b^2_1-M^{12}(b^2_{1,2}+\Gamma^2_{11}b^2_1+\Gamma^2_{22}b^2_1)=0.
\end{equation}
This is the same as what Mockensturm obtained for rectangular, homogeneous ribbons. By also noting that $W''(x_1)$ is small, the boundary conditions on the traction free lateral edges in regions where $W'(x_1)=0$ are:
\begin{equation} 
N^{22}-2b^2_1M^{12}=0,\ \ \ M^{12}_{,1}=0,\ \ \ M^{22}=0.
\end{equation}
Inserting the expressions for fundamental forms, Christoffel symbols and stress resultants calculated in previous sections (see Eqs.~\ref{forms} and \ref{strres}), we obtain a single nonlinear ODE for the lateral stretch of the ribbons. (We assume the dependence of all variables on $x_1$ to be fixed and remove the dependence of the variables on this coordinate in our notation for simplicity.)
\begin{multline}\label{equil4}
\frac{6 f''(x_2)\big(\nu(\lambda_1^2-1+\theta'^2f(x_2)^2)+f'(x_2)^2-1\big)}{f'(x_2)}-\frac{6 \theta'^2f(x_2)\big(\lambda_1^2-1+\theta'^2f(x_2)^2+\nu(f'(x_2)^2-1)\big)}{f'(x_2)}+...\\
+\frac{4H^2\lambda_1^2\theta'^2(1-\nu)f'(x_2)f''(x_2)}{\lambda_1^2+\theta'^2f(x_2)^2}-\frac{H^2\lambda_1^2\theta'^4(1-\nu)f(x_2)f'(x_2)\big(\lambda_1^2+\theta'^2 f(x_2)^2+f'(x_2)^2\big)}{(\lambda_1^2+\theta'^2f(x_2)^2)^2}+...\\
+12f'(x_2)(\theta'^2\nu f(x_2)+f''(x_2))=0.
\end{multline}
The boundary condition $N^{22}-2b^2_1M^{12}=0$ at the lateral edges becomes:
\begin{equation}\label{bcf}
3 (f'(\pm W/2)^2-1+ v (\theta'^2f(\pm W/2)^2+\lambda_1^2-1))+\frac{H^2 \lambda_1^2 \theta'^2 (1-v) f'(\pm W/2)^2}{\theta'^2 f(\pm W/2)^2+\lambda_1^2}=0.
\end{equation}
We set changes of variables $e\equiv (\lambda_1^2-1)/2,\ \eta=H/W$, and define a non-dimensional parameter $T_p = W\theta'$. To proceed with the solution of this differential equation, we note that $T_p$ is small and $e$ and $\eta$ are on the order of $T_p^2$. This determination of order stems from the pre-stretch being very small and from the ribbons having very slender cross-sections, and has been validated numerically for the ribbon geometries we study. We then use a perturbation $f(x_2)=\sum_{I=0}^{\infty}f_{(2I)}(x_2)$ which has a slightly different form compared to what Mockensturm proposed. Inserting this into Eqs.~\ref{equil4} and \ref{bcf} gives us an ODE for each order of the lateral stretches $f_{(2I)}(x_2)$:
\begin{multline}
\mathbf{Zeroth\ order:\ }\\(3 f_{(0)}'(x_2)^2-1) f_{(0)}''(x_2)=0\\
\mathbf{BCs:}\ f_{(0)}'(\pm W/2)^2-1=0,\ \ f_{(0)}=0\\
\mathbf{Solution:}\ f_{(0)}(x_2)=x_2.\\
\end{multline}
\begin{multline}
\mathbf{Second\ order:\ }\\
f_{(2)}''(x_2)+ \frac{\nu T_p^2}{W^2}  x_2=0\\
\mathbf{BCs:\ }f_{(2)}'(\pm W/2)+\frac{T_p^2 \nu}{8}+ e \nu=0,\ \ f_{(2)}=0\\
\mathbf{Solution:}\ f_{(2)}(x_2)=-e \nu x_2-\frac{\nu T_p^2x_2^3}{6W^2}.\\
\end{multline}
\begin{multline}
\mathbf{Fourth\ order:\ }\\
f_{(4)}''(x_2)+\frac{ T_p^4 \left(2 v^2-3\right) }{6W^4} x_2^3-  \frac{e T_p^2}{W^2} x_2=0\\
\textbf{BCs:\ }f_{(4)}'(\pm W/2)+\frac{1}{24} \nu^2 (12 e^2-3 e T_p^2-\frac{T_p^4}{16})=0,\ \ f_{(4)}=0\\
\mathbf{Solution:\ }f_{(4)}(x_2)=\frac{1}{1920}\Bigg(\bigg(15 T_p^4 (\nu^2-1)+240 e T_p^2 (\nu^2-1)-960 e^2 \nu^2\bigg)x_2+\frac{320 e T_p^2}{W^2} x_2^3+\frac{16T_p^4 (3-2 \nu^2)}{W^4} x_2^5\Bigg).\\
\end{multline}

Now that we have calculated $f$ to fourth order, we can insert the function into the expressions for strains (Eq.~\ref{straintensors}) and stresses (Eq.~\ref{strres}) in order to compare predictions from this analytical model to those from the numerical simulations.

\subsubsection{Analytical results and comparison with the numerical ones}

\begin{figure}[!htb]
\centering
\includegraphics[scale=1.1]{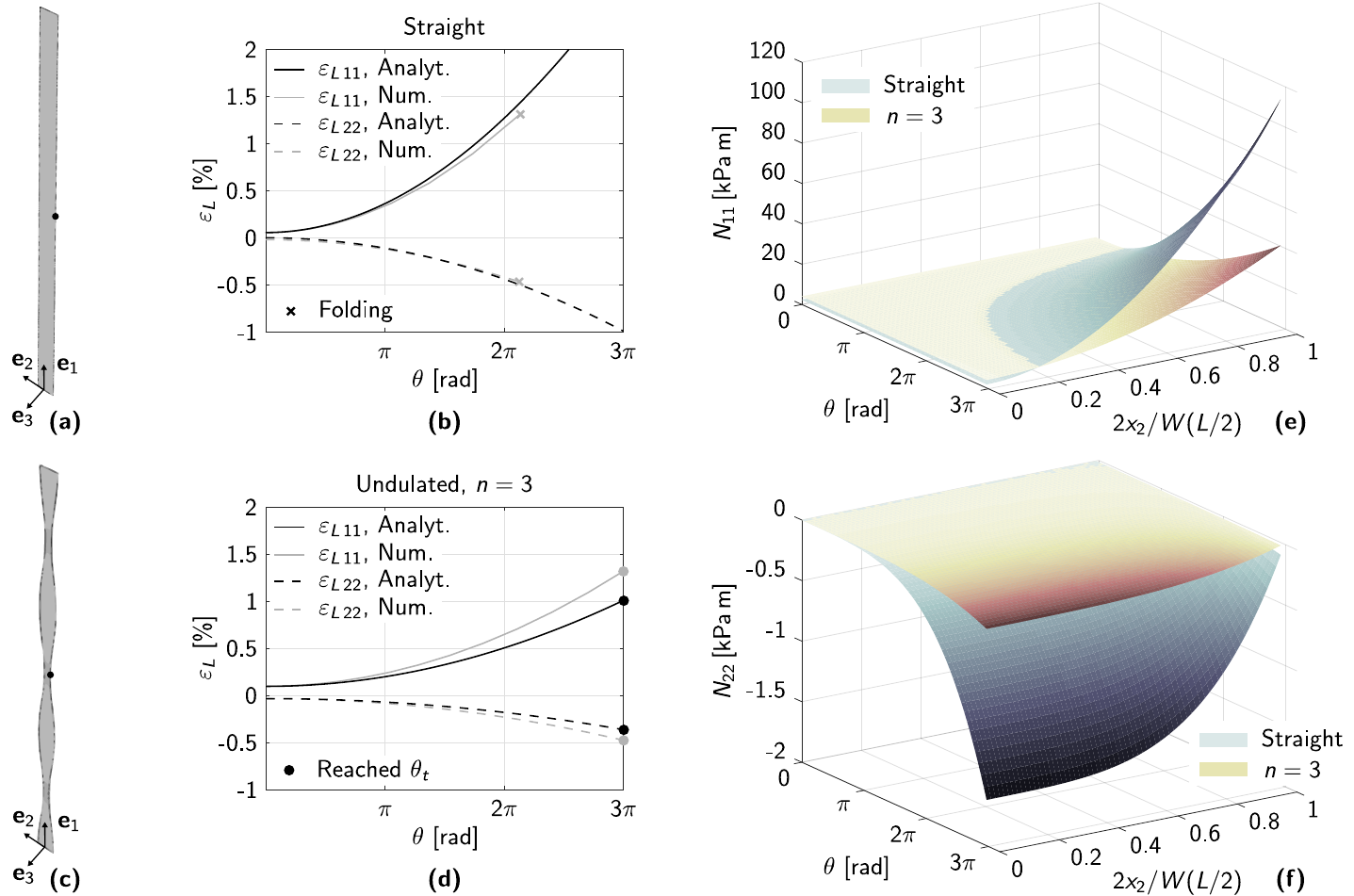}
\caption{Comparison between analytical and numerical results. (a) Schematic diagram of a straight ribbon. (b) Principal logarithmic strains at the free edges of a straight ribbon. (c) Schematic diagram of a ribbon with undulated edges and $A=w/6$. (d) Principal logarithmic strains at the free edges of a ``neck'' region. In (b), (d), the strains are plotted as a function of total relative twist of the clamped edges. (e) Analytical prediction of the axial stress resultant as a function of twist and normalized lateral position in ribbons with straight edges and ribbons with three neck regions. For the undulated-edge ribbon, we plot the stresses at a neck cross-section, where the stresses are greatest. (f) Analytical prediction of the lateral stress resultant as a function of twist and normalized lateral position in ribbons with straight edges and ribbons with three neck regions. The greatest lateral compressive stresses (plotted here) in undulated-edge ribbons emerge in the center of the neck regions.}
\label{f:analystr}
\end{figure}

We first compare the strains predicted by numerical simulations and by this analysis for the case of ribbons with straight edges in Fig.~\ref{f:analystr}(a-b). 
We do the same for a ribbon with three necks and edge undulation amplitude $A=w/6$ (where $w$ is the maximum width of the ribbon) in Fig.~\ref{f:analystr}(c-d). In both of these scenarios, we plot the principal logarithmic strains as a function of clamp twist angle at the region of the ribbon that experiences the greatest principal strains (marked with dots in Fig.~\ref{f:analystr}(a,c)), showing good agreement between the numerical and analytical methods. 
From Fig.~\ref{f:analystr}(b), we can see that the analytics, unlike the numerics, do not capture any self-folding behavior. In Fig.~\ref{f:analystr}(d), we can see that analytics and numerics follow the same trend, especially for the lateral strains. {Discrepancies in this case have to be ascribed to the fact that our theory is only valid for small $A\,n/L$, i.e., for small amplitude to wavelength ratios of the undulation.}

Having shown a consistency between the two methods of analysis, we can now use our analytical results to study the stress evolution within the most vulnerable (narrowest) cross-section. Fig.~\ref{f:analystr}(e-f) shows that the introduction of undulated edges reduces the principal stress resultants considerably. In particular, Fig.~\ref{f:analystr}(f) highlights the emergence of compressive lateral stresses toward the center of the ribbon as twist is increased. These lateral stresses induce buckling at a critical twist. It is clear that the stresses are much greater for ribbons with straight edges than for ribbons of equal length, maximum width and thickness, but with undulated edges. This highlights the benefits of our design strategy when trying to obtain twisted ribbons that do not buckle during twisting. Our analytical model can be extended to the analysis of buckling, as done by Mockensturm \cite{Mockensturm2000}, but this is beyond the scope of this work.

{\subsubsection{Parametric study}
\label{s:param}

We now leverage our analytical model to perform a broader parametric analysis than the one reported in Section~\ref{s:num}. The results of this analysis are reported in Fig.~\ref{f:param}.
\begin{figure}[!htb]
\centering
\includegraphics[scale=1.1]{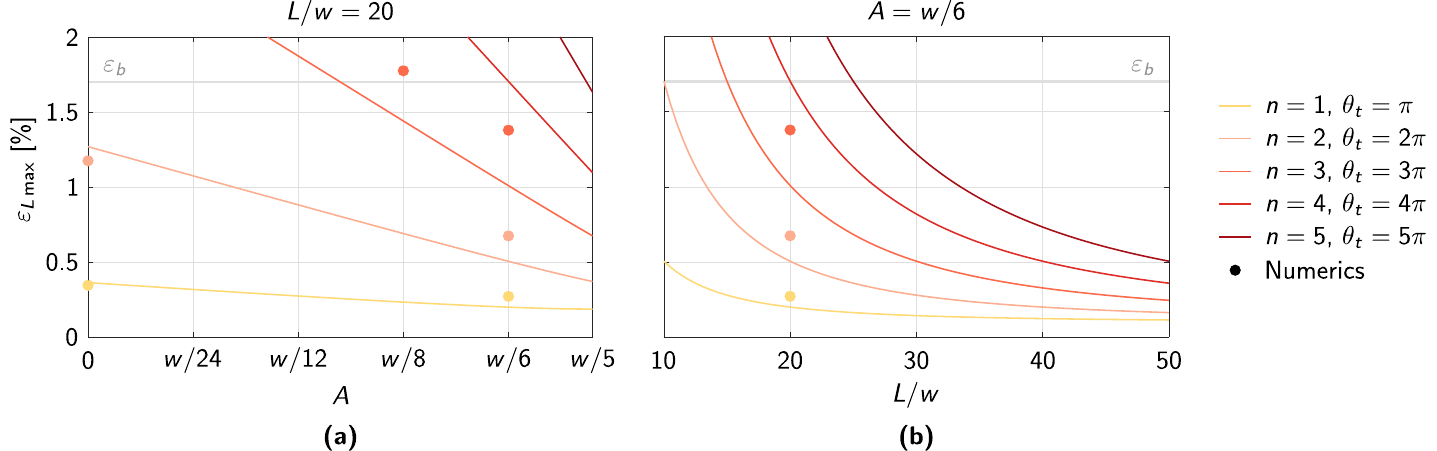}
\caption{Extended parametric analysis. All curves are obtained using our analytical model. The dots are numerical data points and are useful to understand the limitations of the analytical model. (a) Effects of the amplitude of undulation $A$ on the maximum principal strain, for various $n$ and with $L/w=20$ fixed. (b) Effects of $L/w$ for various $n$, with $A=w/6$ fixed.}
\label{f:param}
\end{figure}
First, in Fig.~\ref{f:param}(a), we analyze the effects of $A$ on the maximum principal strain, for various $n$ and keeping the aspect ratio $L/w=20$ fixed. All values correspond to points at the edge of a neck region. We can observe that increasing the undulation amplitude from 0 (straight ribbon) to $w/5$ causes the maximum strains to decrease. We also observe that increasing $n$ causes the level of strain to increase during twisting, noting that $\theta_t$ increases proportionally with $n$. This is consistent with what is shown in Fig.~\ref{f:numstr}(e). The superimposed circular markers follow the same color coding of the analytical lines and represent numerical data points. They allow us to evaluate the performance of the analytical model. We can see once again that the analytical model is more accurate for ribbons where the edge undulation amplitude is much smaller than the undulation wavelength, and tends to significantly underestimate the maximum principal strain for smaller values of $L/(n\,A)$. Also note that the analytical model does not capture whether self-folding occurs before the breaking strain is reached.

In Fig.~\ref{f:param}(b), we analyze the effects of $L/w$, the aspect ratio of the ribbon, on the maximum principal strain. In this case, we fix $A=w/6$. We can see that increasing $L/w$ causes an exponential decrease of the maximum principal strain. While increasing $L/w$ helps delay the onset of failure, it comes at the expense of having compact ribbon geometries.}

\section{From ribbons to structures}
\label{s:struct}

We now have the theoretical and numerical tools to choose geometrical configurations that yield desired shapes upon twisting. Based on previous considerations, we choose ribbons where $n=3$ and $A=w/6$. In this section, we describe the thermoforming process and the setup we designed for twisted ribbon fabrication. We also investigate the potential of single twisted ribbons and assemblies of them as deployable mechanical systems. We do so by analyzing the bending behavior of pre-twisted ribbons and by illustrating prototypes of deployable mechanical systems capable of reversible compaction and deployment cycles.

\subsection{Thermoforming}
\label{s:thermo}

\begin{figure}[!htb]
\centering
\includegraphics[scale=1.1]{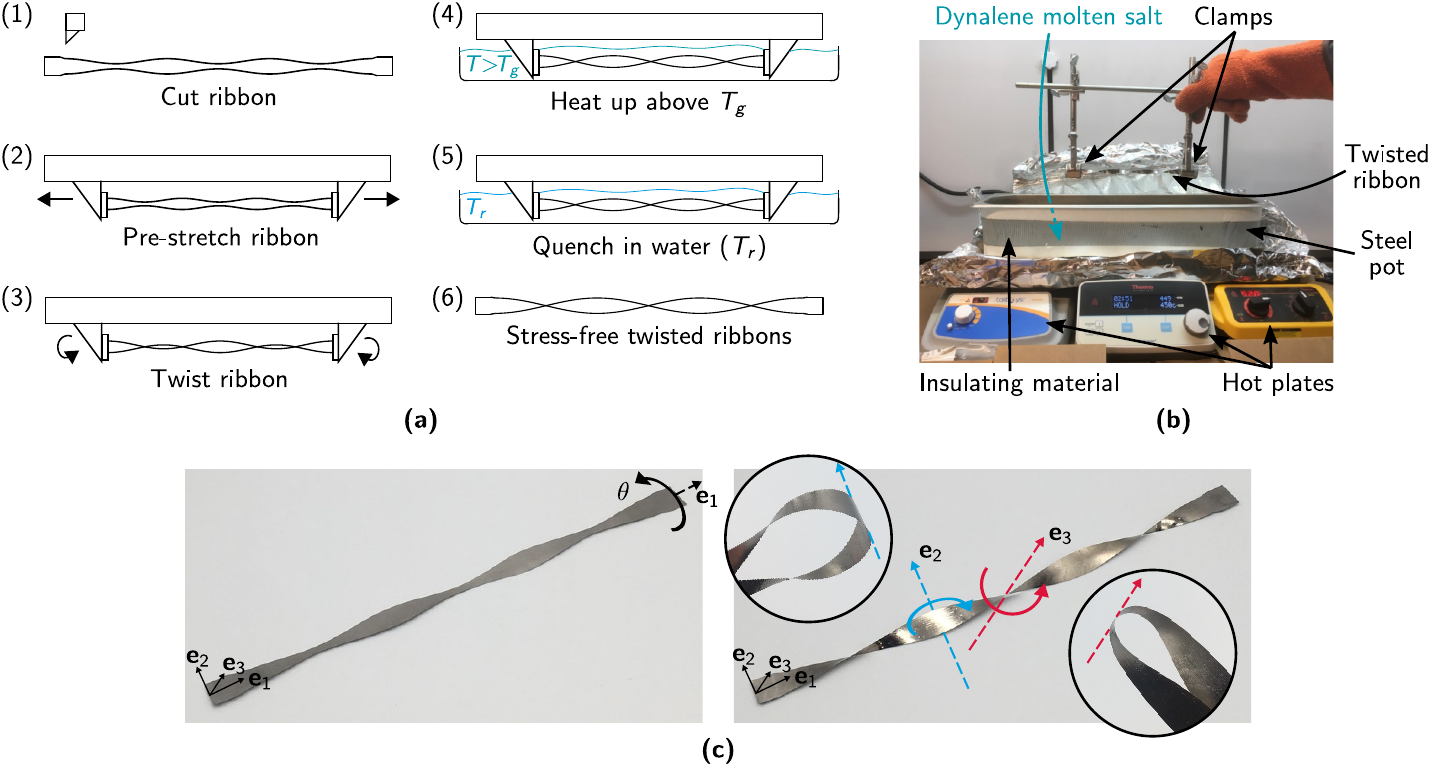}
\caption{BMG ribbon thermoforming. (a) Sketch illustrating the various steps of the thermoforming process, from an initial planar ribbon to a final twisted and stress-free configuration. (b) Thermoforming setup. (c) BMG ribbon before and after thermoforming, with insets illustrating how wide faces and neck regions can be bent about $\textbf{e}_2$ and $\textbf{e}_3$, respectively.}
\label{f:thermo}
\end{figure}

The steps required to thermoform an initially-flat ribbon into a twisted configuration are illustrated in Fig.~\ref{f:thermo}(a). A picture of the fabrication setup is shown in Fig.~\ref{f:thermo}(b).
First, a ribbon is manually cut. Then, we use a custom setup to clamp its supporting tabs (described in Section~\ref{s:design}) and apply a pre-stretch to avoid longitudinal instabilities during the twisting process. The ribbon is then twisted to its target angle and is subsequently immersed in a hot salt bath (Dynalene MS-2). The bath temperature is continuously monitored using a thermocouple and is kept constant at a value that is between the BMG's glass transition $T_g$ and its crystallization temperature $T_x$. This is required for the material to be thermoformable while avoiding the onset of crystallization, which would cause the material to become brittle. For our BMG alloy, we perform Differential Scanning Calorimetery experiments and measure $T_g=370\,\degree{\mathrm C}$ and $T_x=445\,\degree{\mathrm C}$. Thus, we keep the salt bath at $\sim400\,\degree{\mathrm C}$. Our thermoforming protocol consists of immersing a specimen in the bath for 10 seconds, and then quenching it in water at room temperature. This procedure leads to stress-free BMG ribbons that assume the desired twisted shape while preserving the material's elasticity. Note that thermoforming also corrects any curvature induced by the melt-spinning process. Pictures of the 3-neck ribbon before and after thermoforming are shown in Fig.~\ref{f:thermo}(c). The insets in the twisted configuration image show that the neck regions can bend about the $\textbf{e}_3$ axis and the wide faces can bend about $\textbf{e}_2$, as desired. One interesting aspect of these twisted ribbons is that they have an inherent chirality, which is imposed by choosing the twisting direction during fabrication. The case shown in Fig.~\ref{f:thermo}(c), for example, is such that the normal to the surface of the ribbon rotates in a counterclockwise fashion along $\textbf{e}_1$. 

\subsection{Bending behavior of twisted ribbons}
\label{s:bend}

\begin{figure}[!htb]
\centering
\includegraphics[scale=1.1]{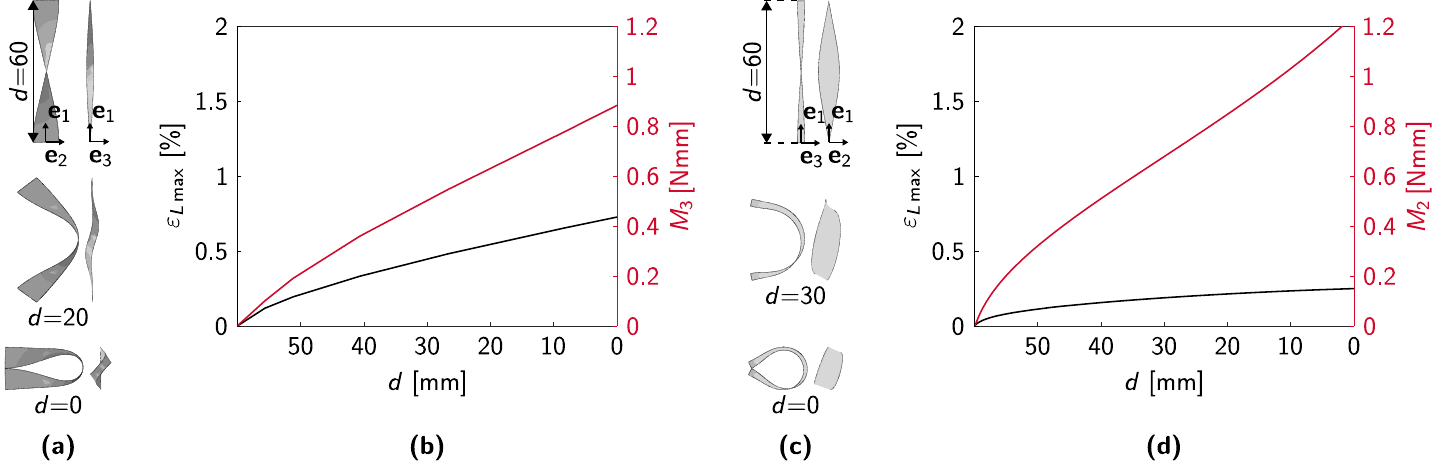}
\caption{Numerical (FE) bending response of different regions of the same twisted ribbon (with $n=3$ and $A=w/6$). (a) Snapshots of the bending deformation of the neck about $\textbf{e}_3$. All dimensions are in mm. (b) Performance of the neck region as a joint, indicating the maximum strain involved and the moment about the rotation axis. (c) Snapshots illustrating how a wide face bends about $\textbf{e}_2$. (d) Bending performance of the wide face.}
\label{f:bend}
\end{figure}

To allow for repeated stowage and deployment of our structures, it is important that bending the necks about $\textbf{e}_3$ and the wide faces about $\textbf{e}_2$ does not produce strains that exceed the breaking strain of the material. To verify that this is the case, we perform bending simulations on the pre-twisted ribbons, {using the same FE model discussed in Section~\ref{s:num}}. To speed up computations, we only consider portions of the selected ribbon geometry. A segment of the pre-twisted stress-free ribbon that includes a single neck and terminates at the midpoints of two consecutive wide faces (thus having length $L/3=60\,\mathrm{mm}$) is illustrated at the top of Fig.~\ref{f:bend}(a).
To simulate bending of a neck about $\textbf{e}_3$, we constrain all points belonging to the top and bottom edges of the ribbon segment to remain in the $\textbf{e}_1$--$\textbf{e}_2$ plane, and we force the two left extremes of the top and bottom edges to displace towards each other ($d$ is the distance between these two points). The bent configurations for $d$ values of $20\,\mathrm{mm}$ and $0\,\mathrm{mm}$ are also shown in Fig.~\ref{f:bend}(a). The evolution of the maximum principal strain in the ribbon and of the moment $M_3$ about $\textbf{e}_3$ as $d$ decreases are illustrated in Fig.~\ref{f:bend}(b). We can see that the strains produced during bending remain below the breaking strain threshold of $1.7\%$. The moment versus displacement plot (\emph{de facto} a moment-angle plot), is obtained by monitoring the resultant force along $\textbf{e}_1$ at the left extreme of the top edge of the ribbon, and by multiplying it by the displacement along $\textbf{e}_2$ of the center of the neck. We can see that the moments are two orders of magnitude smaller compared to hinges that are designed specifically for aerospace applications~\cite{Sakovsky2019}. This implies that small moments are needed to go from the fully-deployed to the stowed configuration and that the deployed structure has limited stiffness. This behavior could be improved either by increasing the structure's dimensions (especially the ribbon's thickness---a choice that would require monitoring strains to prevent failure), or by altering the design to introduce a curvature about $\textbf{e}_1$ that could yield a bistable behavior similar to that displayed by tape-spring hinges. This would require a modification of the thermoforming setup that is not discussed in this article.

We also simulate bending of the wide faces about $\textbf{e}_2$, as illustrated in Fig.~\ref{f:bend}(c,d). In this case, the maximum principal strains achieved are extremely low, since bending a wide face produces low curvatures. However, moments are larger than those in Fig.~\ref{f:bend}(b), indicating that bending a wide face is more difficult than bending a neck. {From Fig.~\ref{f:bend}(a) and (c), we can see that both necks and faces do not behave like perfect planar hinges. Due to the chirality of the ribbons, these regions feature asymmetric bending profiles (e.g., the neck region in Fig.~\ref{f:bend}(a) does not remain symmetric about the $\textbf{e}_1$--$\textbf{e}_2$ plane). This aspect can be leveraged to introduce additional degrees of freedom and enrich the shape-changing capacity of structures made from twisted ribbons.}

\subsection{Tabletop-scale structural prototypes}
\label{s:deploy}

\begin{figure}[!htb]
\centering
\includegraphics[scale=1.1]{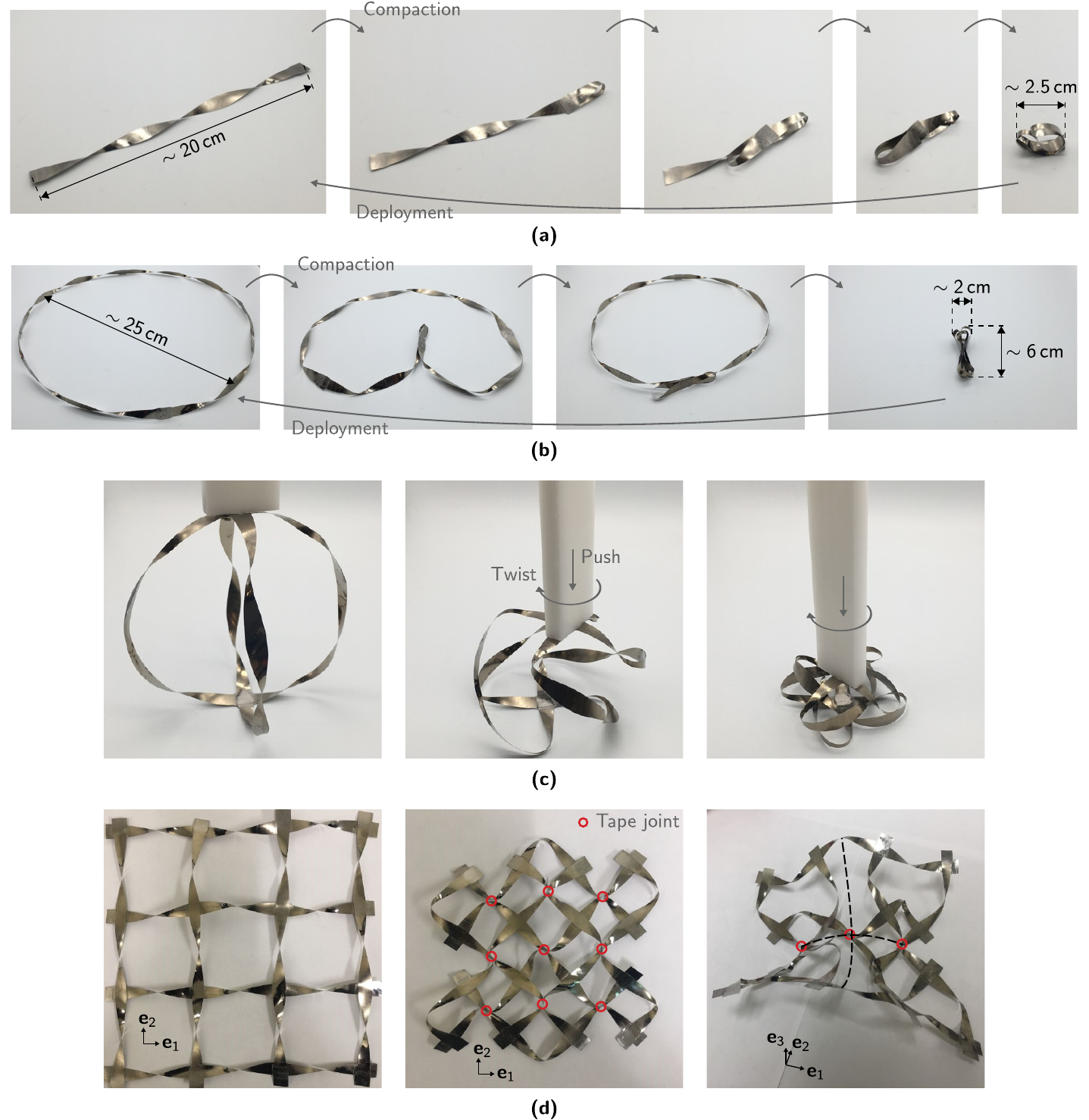}
\caption{Elastic stowage and deployment of twisted ribbons. In all cases, deformations are reversible and do not induce any plastic deformation. Note that we use tape to keep together the stowed configurations for illustration purposes. (a) A twisted ribbon can be compacted by folding it about the necks/hinges, and by finally bending the wide faces. (b) Assembling four ribbons in a circle leads to a ring structure that can be compacted following the same procedure shown in (a). (c) The chirality of the ribbons can be leveraged to create a sphere that can be compacted by applying a twisting load, similarly to Hoberman's Twist-O. {(d) Planar auxetic lattice made of twisted ribbons. The dashed lines highlight the global curvature achievable by taping together selected pairs of necks.}}
\label{f:str}
\end{figure}

Now that we verified that pre-twisted ribbons can bend without breaking, we investigate several deployment-stowage scenarios for single twisted ribbons and assemblies of them. First, we consider a single ribbon, shown in Fig.~\ref{f:str}(a).
In order to compact this one-dimensional structural element, we first fold one of the wide faces onto another, leveraging the hinge-like behavior of one of the necks. We repeat this process sequentially for all wide faces, until we obtain the configuration indicated in the second image from the right. At that point, we bend the stack of wide faces to further compact the system, obtaining the stowed configuration illustrated in the right-most panel. The longest dimension of this compacted ribbon is one order of magnitude smaller than the initial size, highlighting its potential as a deployable system. It is to be noted that the stowed configurations are kept together with double sided tape for illustration purposes, and that the ribbon goes back to the original configuration upon tape removal, owing to the fact that we are not exceeding the breaking strain.

The full potential of these systems as deployable structures can be achieved by combining multiple twisted ribbons in order to create two- and three-dimensional systems. A reliable way of joining multiple ribbons is via spot-welding. By joining four ribbons featuring the same chirality, we obtain the ring shown in Fig.~\ref{f:str}(b), that has an initial diameter of $25\,\mathrm{cm}$. By folding wide faces on top of each other and leveraging the joint-like behavior of the necks as we did in Fig.~\ref{f:str}(a), we can compact the ring, and obtain the final configuration shown in the right-most panel. One can also create three-dimensional structures, as shown in Fig.~\ref{f:str}(c). This sphere is obtained by first creating two rings from ribbons that all have the same chirality. Then, the rings are joined at two couples of wide faces. There are many ways to compact this system, but a particularly interesting one can be achieved by pushing down on the sphere from its top-most point, while simultaneously applying a rotation. This behavior is reminiscent of Hoberman's Twist-O toys, i.e., spheres made by pin-jointed polymeric crosses that can also be compacted by twisting one of their units.  In our case, applying a counterclockwise or clockwise twist produces different stowed configurations owing to the chirality of the ribbons. This further highlights the potential that twisted ribbons have to create structures with many stowage configurations. {Finally, in Fig.~\ref{f:str}(d), we show that twisted ribbons can be used as building blocks for structures with negative Poisson's ratio (a behavior known as auxeticity). From the undeformed lattice, the planar stowed configuration that displays an auxetic behavior (i.e., global shrinkage along $\textbf{e}_1$ and $\textbf{e}_2$) is obtained by taping together alternating pairs of necks, as illustrated in the central panel of the figure. If fewer pairs of necks are joined together, the structure curves into a three-dimensional surface, owing to the chirality and non-planar nature of our hinges. This phenomenon can potentially be tailored by assembling ribbons with different chirality, can be leveraged to create freeform structures from planar precursors, and deserves a separate treatment in future work.}

\section{Conclusions and outlook}
\label{s:concl}

This work represents a first attempt at utilizing thin structural elements made of bulk metallic glass to create compliant deployable structures. We do this by taking initially-flat ribbons and twisting them into structural elements that feature regions that behave as compliant hinges with different preferred bending axes, and assembling these ribbons into more complex three-dimensional structures. Along the way, we use numerical and analytical models to understand the mechanics of twisting and to design ribbons that can be twisted and thermoformed into desired configurations. We also use simulations to verify that pre-twisted ribbons do not fail when they are bent and used as deployable structural systems. Here, we only consider twisted ribbons as building blocks for our structures. However, it could be possible to include ribbons with different deployability attributes, e.g., axial extension, to create structures featuring more complex deformation patterns. 


\section*{Acknowledgments}
This research was carried out at the California Institute of Technology and the Jet Propulsion Laboratory under a contract with the National Aeronautics and Space Administration, and funded through the President's and Director's Fund Program. Reference herein to any specific commercial product, process, or service by trade name, trademark, manufacturer, or otherwise, does not constitute or imply its endorsement by the United States Government or the Jet Propulsion Laboratory, California Institute of Technology. PC and CD acknowledge support from the Foster and Coco Stanback Space Innovation Fund. This work was also supported by a NASA Space Technology Research Fellowship to CM. We thank Michael Mello for helping with material characterization and for fruitful discussions. We thank Basile Audoly and Paolo Ermanni for helpful suggestions, and Brian Ramirez, Sharan Injeti, Hao Zhou, Cristina Naify and Giordano Bellucci for useful discussions.

\appendix
\setcounter{figure}{0} 
\section{Details on the material properties of BMG}
\label{a:mat}

{In this Appendix, we report additional results on the characterization of the BMG chosen in this work, the $\mathrm{Zr_{65}Cu_{17.5}Ni_{10}Al_{7.5}}$ alloy. The engineering stress-strain response of strips of BMG having identical dimensions (up to the precision of our manual cutting process) is shown in Fig.~\ref{f:appmat}.
\begin{figure}[!htb]
\centering
\includegraphics[scale=1.1]{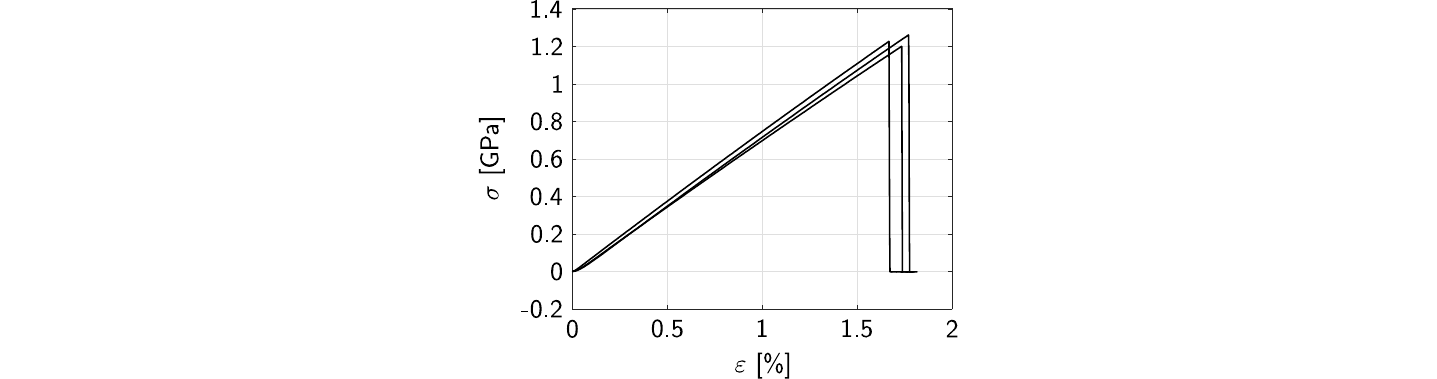}
\caption{Stress-strain response of a BMG strip. Each curve corresponds to an experiment carried out under identical conditions on three specimens.}
\label{f:appmat}
\end{figure}
We can see that all three specimens behave linearly until breaking, and no evidence of plastic deformation is observed. The average breaking stress we obtain from these curves is $\sigma_b\approx1.2\,\mathrm{GPa}$, while the breaking strain (that also represents the elastic strain limit) is $\varepsilon_b\approx1.7\%$. The breaking  strain value is slightly smaller than the nominal one, $\varepsilon_{b}=2\%$~\cite{Zhou2017}, since our experiments are performed in tension and since the melt-spinning process introduces cross-sectional irregularities that can accelerate failure.}

\bibliographystyle{elsarticle-num}

\end{document}